\documentclass[aps,prl,preprint]{revtex4} 
\usepackage{graphicx}

\begin{document}

\title{Tuning the shape of semiconductor microstadium laser}

\author{W. Fang$^1$, G. S. Solomon$^2$, and H. Cao$^1$}
 
\affiliation{$^1$ Department of Physics and Astronomy, Northwestern University, Evanston, IL 60208. \\
$^2$ Solid-State Photonics Laboratory, Stanford University, Stanford, CA  94305; \\ and Atomic Physics Division, NIST, Gaithersburg, MD 20899-8423. } 

\begin{abstract}
We presented a detailed experimental study on lasing in GaAs microstadium with various shapes.  Unlike most deformed microcavities, the lasing threshold varies non-monotonically with the major-to-minor-axis ratio of the stadium. Under spatially uniform optical pumping, the first lasing mode corresponds to a high-quality scar mode consisting of several unstable periodic orbits. By tuning the shape of GaAs stadium, we are able to minimize the lasing threshold. This work demonstrates the possibility of controlling chaotic microcavity laser.  
\end{abstract}

\maketitle

\section{Introduction}

Microcavity lasers are expected to realize an extremely high efficiency and high speed by the volume effect and the spontaneous emission control. Despite rapid developments in semiconductor microcavity lasers, the efficiency is limited by the nonradiative recombination of injected carriers via surface electronic states. As the cavities get smaller, the surface to volume ratio is increased. The resonant modes have more overlap with the cavity boundaries, facilitating the surface recombination. For example, in a circular microdisk cavity, the whispering-gallery (WG) mode is pushed toward the disk boundary as the disk radius decreases. The close proximity of the lasing mode to the cavity boundary where the optical gain is reduced by surface recombination leads to an increase of the lasing threshold. To reduce the detrimental effect of surface recombination on a semiconductor microcavity laser, we propose to use chaotic microcavity to pull the lasing modes away from the boundary \cite{caoLEOS07}. The cavity shape is deformed from circle so that the lasing modes are no longer WG modes. If the modes occupy the interior of a microdisk, the reduction of lasing gain by surface recombination is minimized. Moreover, the carriers in the interior of a microcavity can be utilized for lasing gain. In a circular microdisk, the carriers near the disk center do not contribute to lasing in the WG modes. Such carrier loss is more significant for the case of electrical pumping, because carriers are usually injected to the central region of a microdisk. Another advantage of a chaotic microcavity laser is that it could produce directional output beams. Despite these advantages, the quality factor of a chaotic microcavity is usually low, leading to high lasing threshold. The question we address is whether and how we can minimize the lasing threshold in a chaotic microcavity. In particular, we consider the stadium-shaped microdisk, which is a fully chaotic microcavity.

The stadium billiard has been extensively studied in the fields of classical and quantum chaos \cite{gutzwiller}. The ray mechanics in a stadium billiard exhibits ``full chaos'', i.e., there are no stable periodic orbits. However, a dense set of unstable periodic orbits (UPOs) are embedded in the sea of chaotic orbits. Although the UPOs are found with zero probability in the classical dynamics, in wave (quantum) mechanics they manifest themselves in the eigenstates of the system. There exist extra and unexpected concentrations, the so-called scars, of eigenstate density near UPOs \cite{heller}. Over the past few years, lasing was realized in both scar modes and chaotic modes of semiconductor stadiums with certain major-to-minor-axis ratio \cite{haraPRE,haraPRL}. Highly directional output of laser emission was predicted \cite{schwefel} and confirmed in polymer stadiums \cite{Leb}. In this paper, we demonstrate experimentally that low lasing threshold can be obtained in a semiconductor microstadium by tuning its shape. Contrary to common expectation, modes of such a completely open fully chaotic microcavity may have long lifetime. These special modes are typically scar modes \cite{fang,lee}. When such a mode consists of several UPOs, the interference of partial waves propagating along the constituent orbits may minimize light leakage at certain major-to-minor-axis ratio \cite{fang,fangAPL07}. Thus by controlling the stadium shape, we are able to achieve optimum light confinement in a dielectric microstadium and thus a low lasing threshold.

The paper is organized as follows. Section II describes the fabrication and characterization of GaAs microstadium lasers with various shapes. In Section III, a detailed theoretical analysis of lasing modes in microstadiums is presented. Section IV contains a brief discussion and conclusion.    

\section{Fabrication and characterization of microstadium lasers}

The sample was grown on GaAs substrate by molecular beam epitaxy. The layer structure consists of  500nm AlGaAs and 200nm GaAs. In the middle of the GaAs layer there is a narrow  InAs quantum well (QW) which serves as the gain medium. The AlGaAs layer separates the GaAs layer from the substrate. Its lower refractive index leads to the formation of a slab waveguide in the GaAs layer. Stadium-shaped cylinders were fabricated by photolithography and wet chemical etching. The major-to-minor-axis ratio of the stadiums was varied over a wide range while the stadium area remains nearly constant. For each fabricated stadium, we extracted its actual size and shape from the scanning electron microscope (SEM) image.  Figure 1 shows a top-view SEM image of a GaAs stadium.  The radius of the half circles $r = 2.69\mu$m, the length of the straight segments connecting the two half circles $2a = 8.12\mu$m. The deformation of the stadium is defined as $\epsilon \equiv a/r$. 

To study their lasing properties, the stadium microcavities were cooled to 10K in a cryostat, and optically pumped by a mode-locked Ti-sapphire laser at 790nm. The pump beam was focused by an objective lens onto a single stadium. The emission was collected by the same lens, and sent to a spectrometer with a liquid nitrogen cooled charge coupled device (CCD) array detector. Lasing was realized in most stadiums with $\epsilon$ ranging from 0.4 to 2.2 and area $\sim 70\mu$m$^2$. Figure 2(a) shows the emission spectra of twelve stadiums slightly above their lasing thresholds so that we mainly see the first lasing mode. As $\epsilon$ increases, the first lasing mode jumps back and forth within the gain spectrum of InAs QW. It is not always located near the peak of gain spectrum.  At some deformation, e.g. $\epsilon$ = 0.94, the first lasing mode is far from the gain maximum at $\lambda \sim$ 857nm. This phenomenon is not caused by lack of cavity modes near the maximum of gain spectrum. A few small and broad peaks in the emission spectrum between 847nm and 857nm represent the cavity modes. These modes experiences higher gain than the lasing mode at $\lambda \approx 847$nm. The only reason they do not lase is their quality ($Q$) factors are low. This result indicates the lasing modes, especially the first one, must be high-quality modes. However, when the lasing mode is away from the maximum of gain spectrum, the relatively low optical gain at the lasing frequency results in high lasing threshold. This is confirmed in Fig. 2(b), which shows the lasing threshold strongly depends on the spectral distance between the first lasing mode and the maximum of gain spectrum. Contrary to the common expectation, the lasing threshold in a microstadium does not increase monotonously with the deformation. For example, the lasing thresholds in stadiums of $\epsilon$ = 0.7 and 2.2 are nearly the same despite of their dramatically different deformations. Such behavior of microstadium laser is very different from the elliptical or quadrupolar-shaped microcavity laser whose threshold usually rises continuously with increasing major-to-minor-axis ratio when the cavity area is fixed. 

To investigate individual lasing modes in microstadiums, we used a narrow band pass filter to select one lasing mode and took its near-field image with a CCD camera. Figure 3 shows the measurement result of a stadium with $\epsilon = 1.9$. The solid curve is the emission spectrum when the narrow bandpass filter is tuned to the first lasing mode at $\lambda$ = 842nm. The inset A is the near-field image taken simultaneously. It exhibits six bright spots on the curved part of the stadium boundary. We believe these six spots represent the positions of major escape of laser light from the stadium. They can be seen from the top because of optical scattering at the boundary. However, the scattering inside the stadium is so weak that the spatial intensity distribution of lasing mode across the stadium could not be observed from the top. By tuning the bandpass filter away from cavity resonances (the spectrum plotted by dashed line in Fig. 3), we obtained the near-field image of amplified spontaneous emission (ASE) shown as the inset B of Fig. 3. The virtually constant intensity along the curved boundary suggests the ASE leaves the stadium mainly through the boundary of half circles instead of the straight segments. The clear difference between the near-field images of lasing mode and ASE confirms the bright spots in the former are from the laser emission. 

Figure 4 shows the measurement result of another stadium with $\epsilon = 1.51$. Although the ASE image looks similar, the lasing mode image is quite different from that in the stadium with $\epsilon = 1.9$. The first lasing mode at $\lambda = 857$nm exhibits four bright spots on the curved part of the stadium boundary. It indicates the first lasing modes in these two stadiums have distinct spatial profiles. 

\section{Theoretical analysis of lasing modes} 

To understand the above experimental results, we simulated lasing in GaAs microstadiums. The polarization measurement of laser emission from the side wall of stadiums revealed the lasing modes are transverse electric (TE) polarized, namely, the electric field is parallel to the top surface of the stadium. From the calculated spatial profile of TE wave guided in the GaAs layer, we obtained the effective index of refraction $n_{eff} \simeq 3.3$. The exact size and shape of the fabricated stadiums were extracted from the digitized SEM images. Using the finite-difference time-domain (FDTD) method, we solved the Maxwell's equations for electromagnetic (EM) field inside and outside a two-dimensional (2D) stadium of refractive index $n_{eff}$ together with the four-level rate equations for electronic populations in the InAs QW \cite{nagra}. Light exiting the stadium into the surrounding air was absorbed by the uniaxial perfectly matched layers. The external pumping rate for electronic populations was assumed uniform across the stadium, similar to the experimental situation. We gradually increased the pumping rate until one mode started lasing. Fourier transform of the EM field gave the frequency of the first lasing mode. 

Figure 5(a) shows the calculated intensity distribution of the first lasing mode at $\lambda$ = 826nm in the stadium with $\epsilon$ = 1.9 and area $\simeq 70 \mu$m$^2$. The pumping rate is slightly above the lasing threshold. For comparison, we also calculated the high-quality modes in the passive stadium (without optical gain). Details of our numerical method can be found in Ref.\cite{fang}. As shown in Fig. 5(a), the spatial profile of the lasing mode is nearly identical to that of the cold-cavity mode. This result illustrates the nonlinear effect on the lasing mode \cite{haraPRL2} is insignificant when the pumping rate is not far above the lasing threshold.  

The intensity of light escaping through the stadium boundary can be approximated by the intensity just outside the boundary. From the calculated lasing mode profile, we extracted the intensity of light about 100nm outside the stadium boundary. To account for the finite spatial resolution in our experiment, the output intensity distribution along the stadium boundary was convoluted with the resolution function of our imaging system. The final result [dashed curve in Fig. 6 (a)] agrees well with the measured intensity along the stadium boundary (solid curve), especially it reproduces the positions of the six bright spots in the near-field image of the lasing mode. Thus we conclude the first lasing mode observed experimentally in the stadium of $\epsilon$ = 1.9 corresponds to the calculated high-$Q$ mode at $\lambda$ = 826nm. The slight difference in wavelength (less than 2\%) is within the experimental error of determining the refractive index of GaAs and AlGaAs at low temperature. 

To determine the classical ray trajectories corresponding to the lasing mode, we obtained the quantum Poincar\'e sections of their wavefunctions. Figure 5(b) is the Husimi phase-space projection of the lasing mode in Fig. 5(a), calculated from its electric field at the stadium boundary. It reveals the lasing mode is a scar mode, and it consists mainly of two types of UPOs plotted in the inset of Fig. 5(b). Both orbits are above the critical line for total internal reflection, suggesting the photon dwell time in the cavity is long. However, the mode frequency is away from the peak of gain spectrum, thus the lasing threshold is relatively high. 

The escape of ASE from a stadium is simulated by classical ray tracing in real space, as the interference effect can be neglected due to lack of coherence in ASE. We traced $1.2 \times 10^5$ rays inside the stadium. The initial distribution of rays is uniform in real space and in angular direction. Each ray is initially given the amplitude of one. At each reflection from the boundary, the amplitude is reduced according to the Fresnel formula. The outgoing amplitude is recorded at the position of escape, and reflected ray is followed until its amplitude falls below $10^{-4}$. Figure 6(b) shows the intensity distribution of output rays along the stadium boundary (dashed curve). The ray-tracing result agrees with the ASE intensity distribution obtained from the near-field image (solid curve).

We repeat the above calculations for the stadium with $\epsilon$ = 1.51. The results are presented in Figs. 7 and 8. The spatial profile of the first lasing mode is almost the same as that of a high-$Q$ mode in the passive stadium [Fig. 7(a)]. The calculated output intensity distribution along the stadium boundary reproduces the four bright spots observed experimentally [Fig. 8(a)]. Figure 7(b) shows that the first lasing mode in the stadium with $\epsilon$ = 1.51 consists of three different types of UPOs. We calculated the quality factor of this mode in passive stadium as we varied the deformation $\epsilon$ around 1.51. Its $Q$ value first increases then decreases as $\epsilon$ increases, leading to a local maximum at $\epsilon$ = 1.515 [Fig. 9]. Such variation of quality factor is attributed to interference of waves propagating along the constituent UPOs \cite{fang,fukuJLT,fukuPRA}. The interference effect depends on the relative phase of waves traveling in different orbits. The phase delay along each orbit changes with the orbit length as $\epsilon$ varies. At some particular deformation, constructive interference may minimize light leakage out of the cavity, thus maximize the quality factor. Since the actual deformation $\epsilon=1.51$ is nearly identical (within 0.3\%) to the optimum deformation ($\epsilon=1.515$), the mode is almost at the maximum of its quality factor. Furthermore, its frequency is close to the peak of gain spectrum. Thus the lasing threshold is minimized, as shown in Fig. 2(b). 

\section{Discussion and Conclusion}

We have simulated lasing in fabricated microstadiums with various deformations. By comparing the simulation results with the experimental data, we find the first lasing modes always correspond to high-quality scar modes of the passive cavities. This is because the gain spectrum of InAs QW is broad enough to cover some of these modes. Note that not all the scar modes have high quality or long lifetime. Nevertheless the chaotic modes always have short lifetime, because their relatively uniform distributions in the phase space facilitate the refractive escape of light from the stadium.  If the gain spectrum is too narrow to contain any high-$Q$ scar modes, lasing may occur in low-$Q$ chaotic modes lying within the gain spectrum \cite{haraPRL} but with much higher threshold. 

One surprising result of our experiment is that the lasing threshold in the stadium-shaped GaAs cylinder is usually lower than that in the circular cylinder with the same area. We attribute it to the reduction of surface recombination in the GaAs stadium as the lasing modes occupy the interior of the stadium. This result illustrates an advantage of the microstadium laser. 

Another advantage of the microstadium laser is that the $Q$-spoiling is effectively stopped at large deformation \cite{fang}. High-$Q$ modes exist at large $\epsilon$, as a result of the non-monotonic change of quality factors of some multi-orbit scar modes with deformation. Indeed, we observed different types of high-$Q$ scar modes consisting of several UPOs at various deformations in our simulation. This observation is supported by our experiment result that the lasing threshold at large deformation $\epsilon = 2.2$ is nearly the same as that at small deformation $\epsilon = 0.7$. Since the lasing wavelengths are nearly the same, the optical gain experienced by the first lasing modes in these two stadiums of same area is almost identical. Therefore, the almost same lasing threshold implies the quality factor of the first lasing mode in the stadium with $\epsilon = 2.2$ is nearly identical to that with $\epsilon = 0.7$.  The effective halting of $Q$ spoiling does not exist in the elliptical cavity (an integral system) or the quadrupolar cavity (a partially chaotic system) \cite{nockel,narimanov}. Those systems exhibit continuous $Q$ spoiling with increasing deformation while the cavity area is fixed \cite{fang}. Thus a global increase of lasing threshold is expected when the major-to-minor-axis ratio of dielectric ellipse or quadrupole increases. 

In summary, we demonstrated experimentally that lasing in a semiconductor microstadium can be optimized by controlling its shape. When the optical gain spectrum is broad enough, the first lasing mode in a GaAs microstadium with large deformation is typically a high-quality scar mode consisting of several UPOs. The interference of waves propagating along the constituent orbits results in an optimum deformation at which the quality factor reaches the maximum. By tuning the stadium shape to the optimum deformation, we not only optimize light confinement in the stadium but also extract the maximum gain by aligning the mode frequency to the peak of gain spectrum. The simultaneous realization of the lowest cavity loss and the highest optical gain leads to minimum lasing threshold of a microstadium laser under spatially uniform optical pumping. As the dielectric microstadium represents a completely open fully chaotic cavity, this work opens the door to control chaotic microcavity lasers by tailoring its shape.

We acknowledge Prof. Peter Braun and Dr. Gabriel Carlo for stimulating discussions. This work is supported by NIST under the Grant No. 70NANB6H6162 and by NSF/MRSEC under the Grant No. DMR-00706097.


\begin{figure}
\includegraphics[width=12cm]{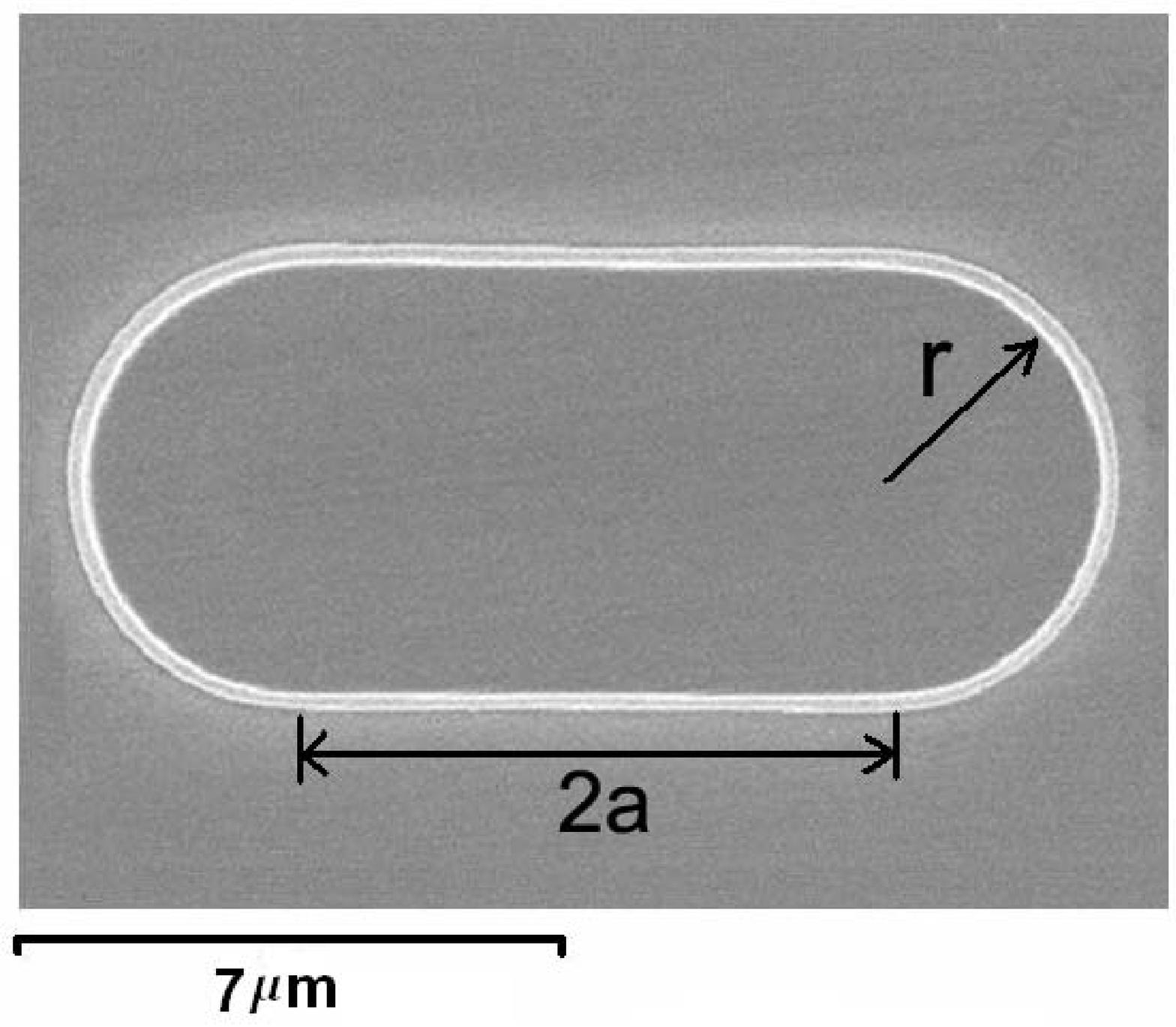}
\caption{\label{fig1}Top-view SEM image of a GaAs stadium. $r = 2.69\mu$m, $2a = 8.12\mu$m, $\epsilon = 1.51$.}
\end{figure}

\begin{figure}
\includegraphics[width=12cm]{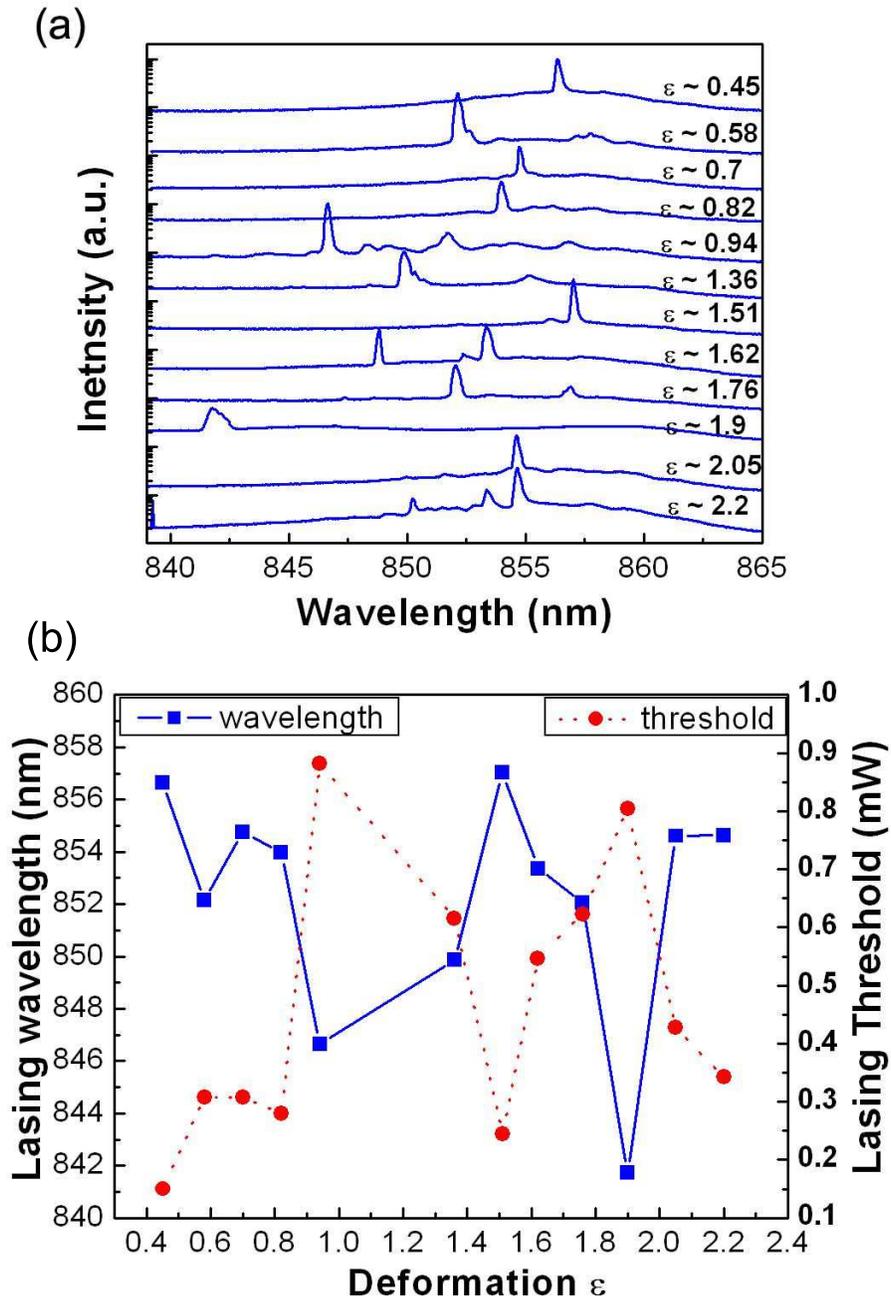}
\caption{\label{fig2}(a) Lasing spectra from twelve GaAs stadiums with different deformations. (b) Wavelength and threshold of the first lasing mode as a function of the deformation.}
\end{figure}

\begin{figure}
\includegraphics[width=12cm]{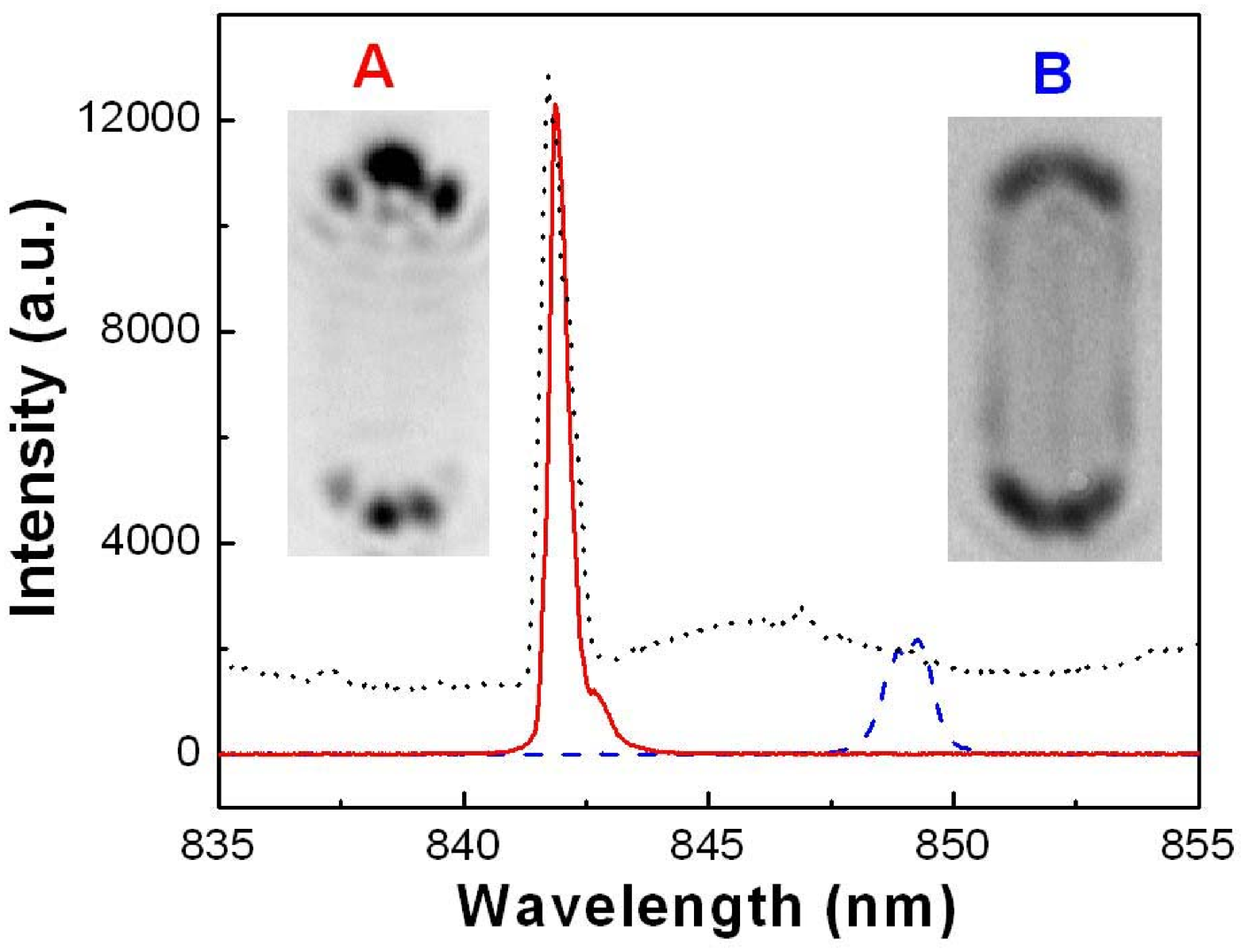}
\caption{\label{fig3}The dotted curve is the lasing spectrum from a GaAs stadium with $\epsilon$ =1.9. A bandpass filter of 1nm bandwidth selects the first lasing mode at 842nm (solid curve), and the inset A is the corresponding near-field image. The dashed curve is the spectrum when the bandpass filter is tuned away from cavity resonances, the corresponding near-field image of ASE is shown in the inset B. }
\end{figure}

\begin{figure}
\includegraphics[width=12cm]{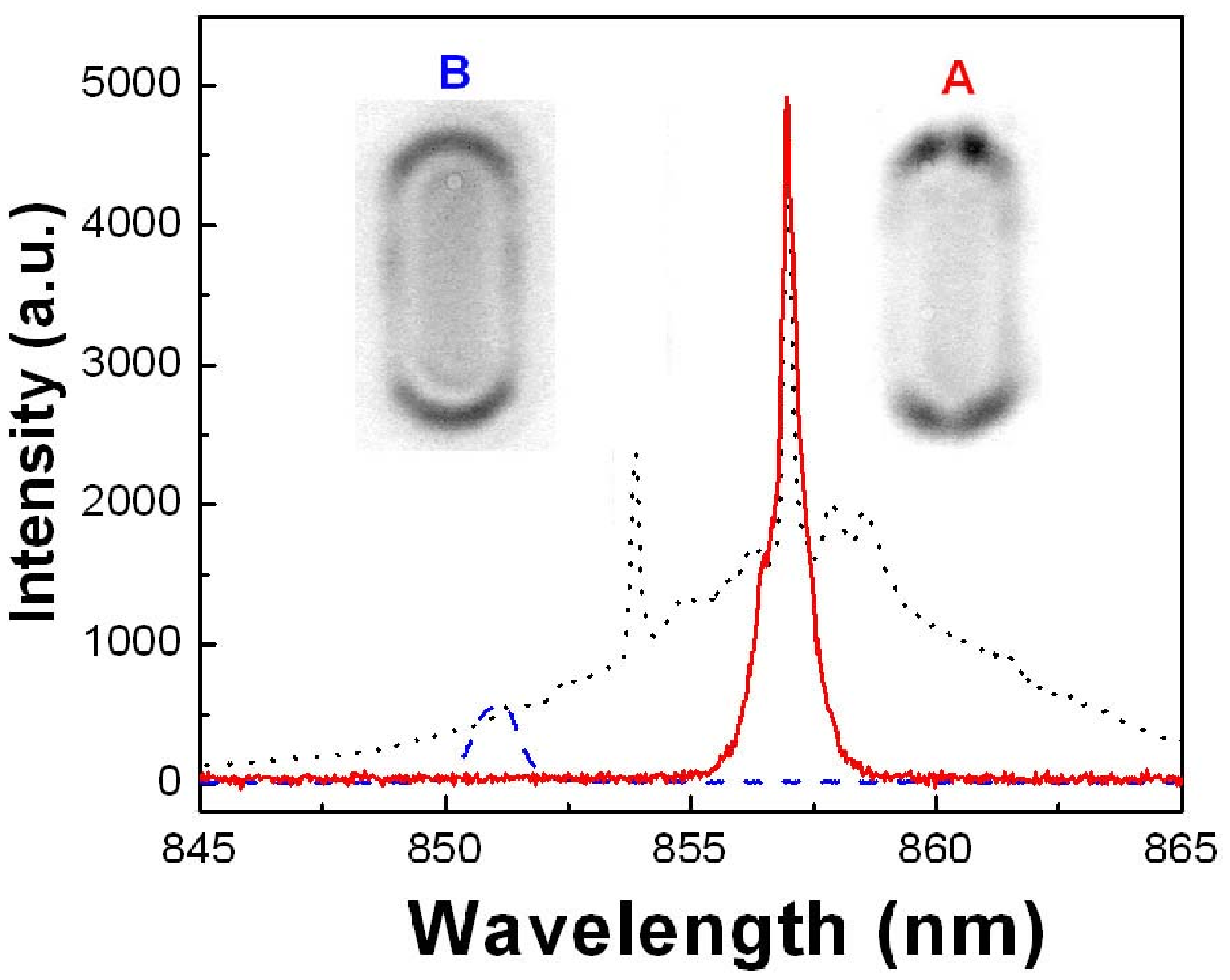}
\caption{\label{fig4}The dotted curve is the lasing spectrum from a GaAs stadium with $\epsilon$ =1.51. A bandpass filter of 1nm bandwidth selects the first lasing mode at 857nm (solid curve), and the inset A is the corresponding near-field image. The dashed curve is the spectrum when the bandpass filter is tuned away from cavity resonances, the corresponding near-field image of ASE is shown in the inset B.}
\end{figure}

\begin{figure}
\includegraphics[width=12cm]{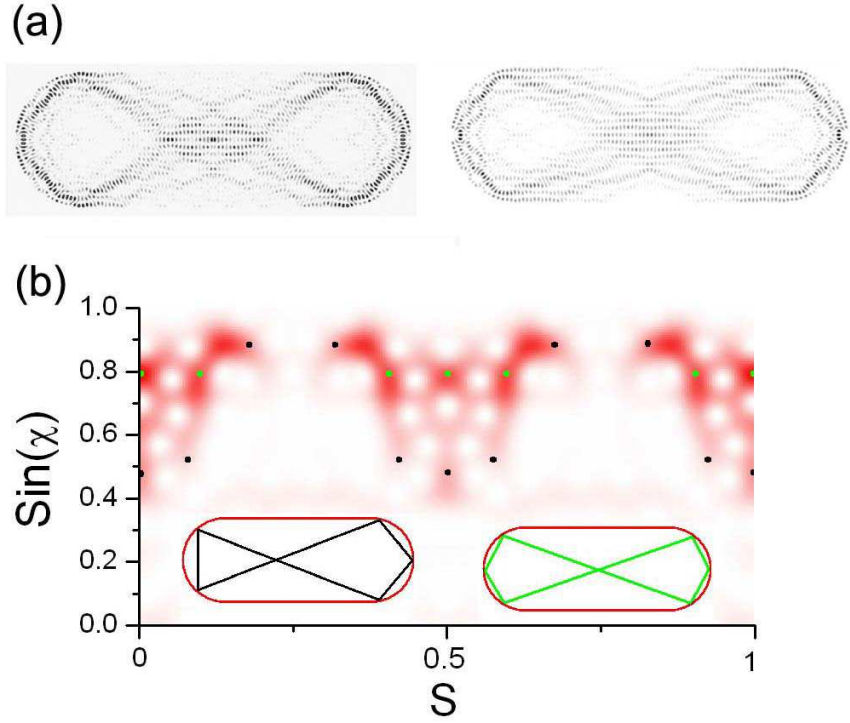}
\caption{\label{fig5}(a) Calculated intensity distribution of the first lasing mode in a stadium with $\epsilon$ = 1.9 (left), and the corresponding mode in the passive stadium without gain (right). Both modes have the wavelength 826nm. (b) Husimi phase space projection of the mode in (a). The horizontal coordinate $S$ represents the length along the stadium boundary from the rightmost point, normalized by the stadium perimeter. The vertical axis corresponds to ${\rm sin} \chi$, where $\chi$ is the incident angle on the stadium boundary. Solid squares and circles mark the positions of two different types of UPOs shown in the inset.}
\end{figure}

\begin{figure}
\includegraphics[width=12cm]{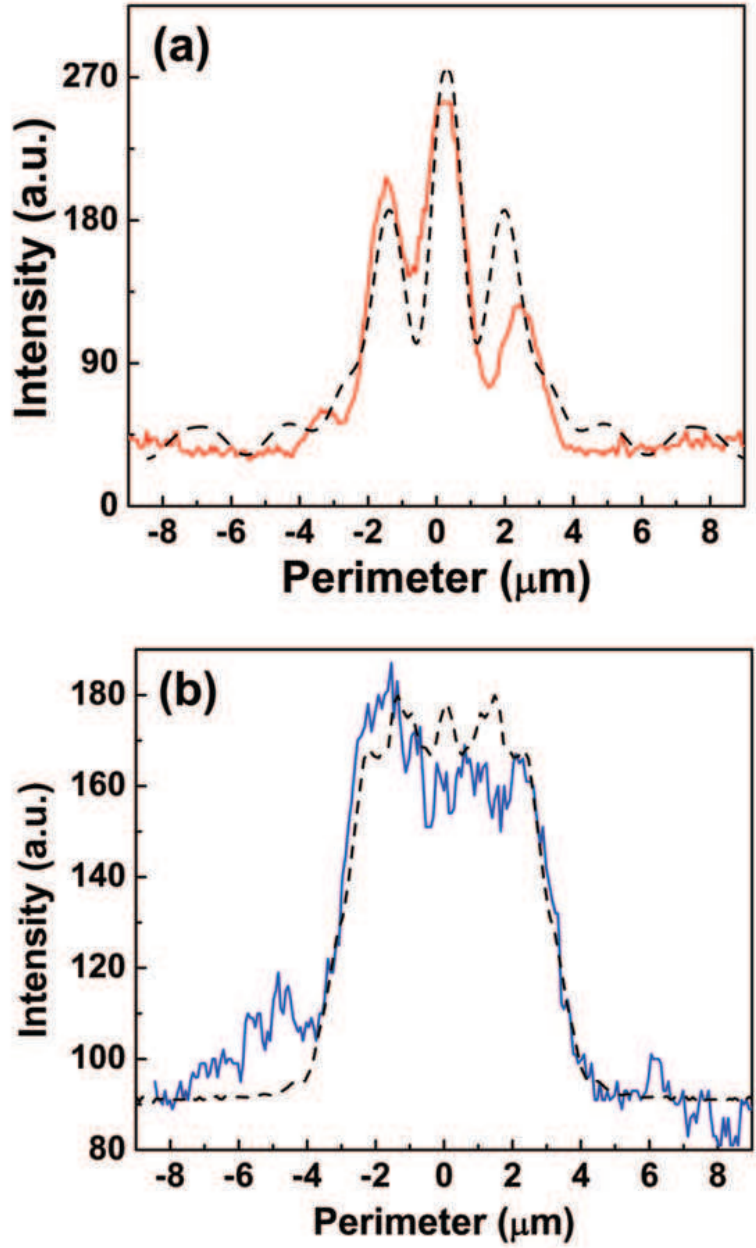}
\caption{\label{fig6}Output intensity of laser emission (a) and ASE (b) along the boundary of a GaAs stadium with $\epsilon$ = 1.9 and area $\simeq 70\mu$m$^2$. The range of the horizontal coordinate is half of the stadium boundary, from the center of one straight segment to the other. The solid curves are the experimental results extracted from the near-field images of the lasing mode and ASE in the insets of Fig. 3. The dashed curves are the numerical simulation results obtained with the FDTD method (a) and real-space ray tracing (b).}
\end{figure}

\begin{figure}
\includegraphics[width=12cm]{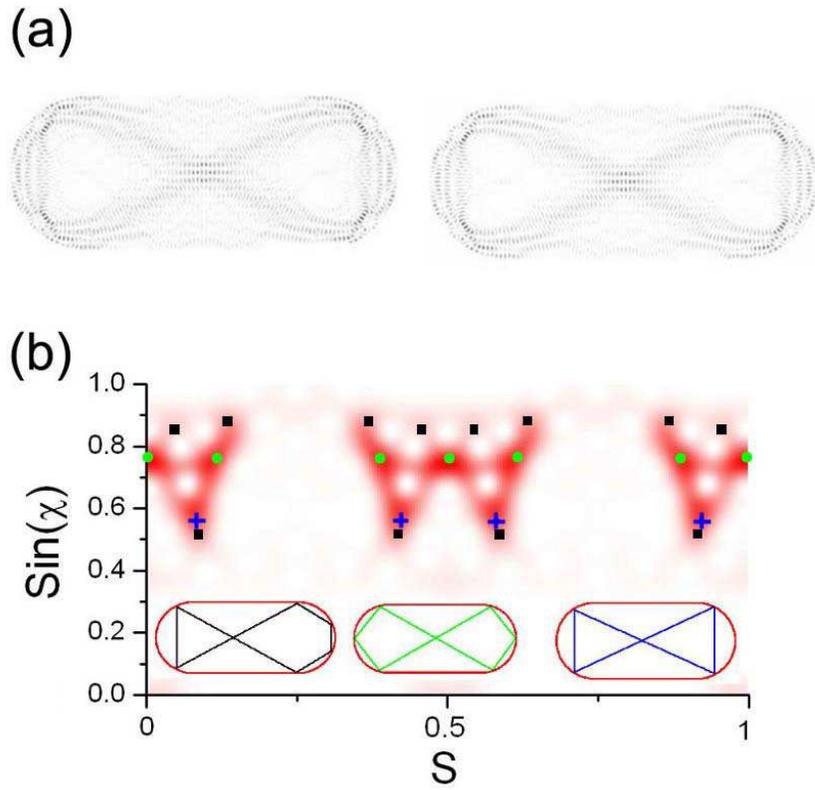}
\caption{\label{fig7}(a) Calculated intensity distribution of the first lasing mode in a stadium with $\epsilon$ = 1.51 (left), and the corresponding mode in the passive stadium without gain (right). Both modes have the wavelength 850.7nm. (b) Husimi phase space projection of the mode in (a). The squares, dots and crosses mark the positions of three different types of UPOs shown in the inset.}
\end{figure}

\begin{figure}
\includegraphics[width=12cm]{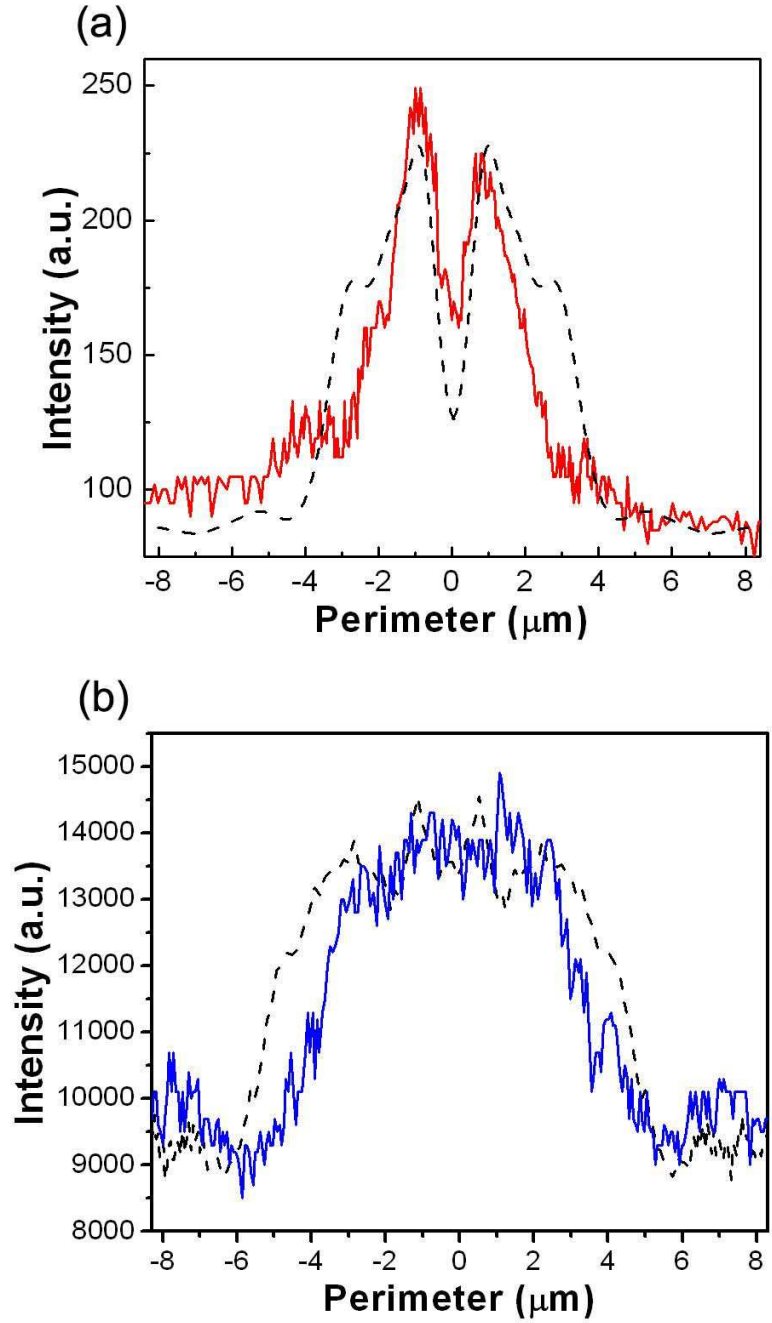}
\caption{\label{fig8}Output intensity of laser emission (a) and ASE (b) along the boundary of a GaAs stadium with $\epsilon$ = 1.51. The solid curves are the experimental results extracted from the near-field images of the lasing mode and ASE in the insets of Fig. 4. The dashed curves are the numerical simulation results obtained with the FDTD method (a) and real-space ray tracing (b).}
\end{figure}

\begin{figure}
\includegraphics[width=12cm]{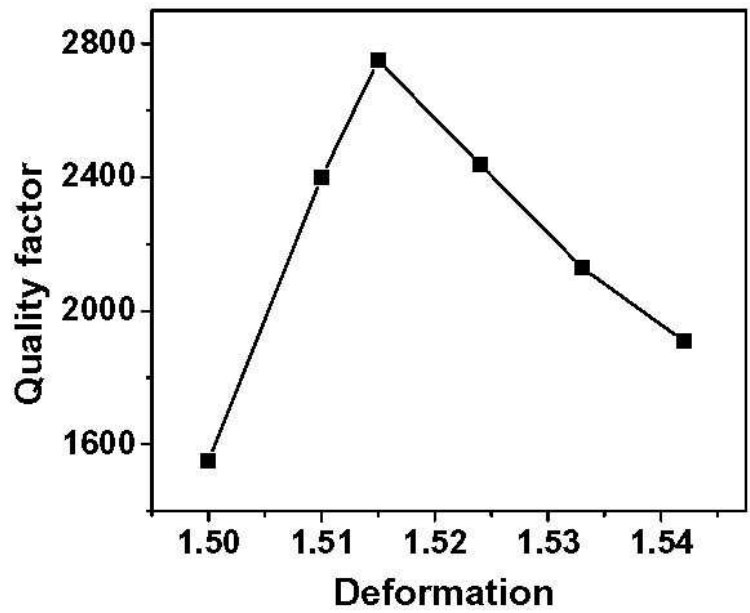}
\caption{\label{fig9}Calculated $Q$-factor of the mode in Fig. 8(a) as a function of $\epsilon$.}
\end{figure}

\end{document}